\documentclass[amssymb,prb,twocolumn,floats,amsmath]{revtex4}

\usepackage[T1]{fontenc}
\usepackage{bm}
\usepackage{graphicx}
\usepackage{amssymb}
\usepackage{amsfonts}
\usepackage{amsmath}
\usepackage{epstopdf}

\begin{document}

\title{Strong coupling and polariton lasing in Te based microcavities embedding (Cd,Zn)Te quantum wells}

\author{J.-G.~Rousset}
\email{j-g.rousset@fuw.edu.pl}
\author{B.~Pi\k{e}tka}
\author{M.~Kr\'{o}l}
\author{R.~Mirek}
\author{K.~Lekenta}
\author{J.~Szczytko}
\author{J.~Borysiuk}
\author{J.~Suf\mbox{}fczy\'nski}
\author{T.~Kazimierczuk}
\author{M.~Goryca}
\author{T.~Smole\'nski}
\author{P.~Kossacki}
\author{M.~Nawrocki}
\author{W.~Pacuski}

\affiliation{Institute of Experimental Physics, Faculty of Physics,
University of Warsaw, ul. Pasteura 5, PL-02-093 Warszawa, Poland}


\begin{abstract}
We report on properties of an optical microcavity based on (Cd,Zn,Mg)Te layers and embedding (Cd,Zn)Te quantum wells. The key point of the structure design is the lattice matching of the whole structure to MgTe, which eliminates the internal strain and allows one to embed an arbitrary number of unstrained quantum wells in the microcavity. We evidence the strong light-matter coupling regime already for the structure containing a single quantum well. Embedding four unstrained quantum wells results in further enhancement of the exciton-photon coupling and the polariton lasing in the strong coupling regime.
\end{abstract}

\maketitle

\section{Introduction}

Observed for the first time in 1992 microcavity polaritons\cite{Weisbuch_PRL92, Savona_PRB94} -- hybrid exciton-photon quasiparticles arise when quantum well (QW) excitons strongly couple with an electromagnetic field in a microcavity. Polaritons are composite bosons and as such they are subject to Bose-Einstein  statistics.\cite{Hulin_PRL80, Kavokin_MC2007} In particular, this results in the increase in the rate of scattering to the final state which is proportional to $1 + N$, where $N$ is the population of the final state \cite{Miesner_Science98} (so-called final state stimulation). This phenomenon underlies the occurrence of Bose-Einstein condensation of  microcavity polaritons\cite{Deng_science2002,Kasprzak_nature2006} and the operation of the so called polariton laser.\cite{Imamoglu_PRA96} In polariton laser, unlike photon laser, light is emitted spontaneously. However, since the emission results from the decay of identical polaritons, transferring their energy to the photons which escape from the crystal, the light is coherent and monochromatic.\cite{Bajoni_PRB2007, Bajoni_PRL2008, Kamman_NJP2012}
Extensive research on microcavity polaritons from its beginning includes both: fundamental aspect in the field of many-body theories and cavity quantum electrodynamics,\cite{Weisbuch_JofLUM2000} as well as polariton laser applications such as new sources of coherent and non classical light with extremely low energy threshold,\cite{Deng_PNAS2003} high-speed polarization switches \cite{Amo_NatPhot2010} or sources of terahertz radiation.\cite{Kavokin_APL2010} All of these studies and possible applications require the design and growth of structures that meet some specific criteria. One of them is a high oscillator strength of the exciton transition, a prerequisite for the light-matter strong coupling regime. Also, the exciton binding energy should be large enough to ensure the operation at reasonably high temperatures and to damp dissociation of the polaritons under a strong excitation, when the density of excitons is high. Finally, we note that strength of exciton-photon coupling increases with a number of QWs embedded into the microcavity, since excited excitons are distributed over a greater volume, what lowers their local density.

Here, we present structures that meet all of these requirements. They are made of II-VI compounds, which ensures relatively large exciton binding energies. Our design of distributed Bragg reflectors (DBR) and the cavities materials fully eliminates any harmful stresses, enabling several identical QWs to be precisely located within the cavity. In addition, it guarantees a high reproducibility of the structures, which is particularly important in studies requiring the tuning of optical properties. It should be stressed that in contrast to previous designs of Te-based microcavities,\cite{Ulmer_SLandMicrostruc97, Sadowski_PRB1997, Cubian_PRB2003, Brunetti_SLandMicrostruc2007} in the growth process of the presented basic DBR structures no magnetic elements are used. This enables their well controlled placement inside the structures, for example in QWs, which makes them especially suitable for future research on paramagnetic impurities and the influence of magnetic field on the exciton-polaritons effects. The microcavity structures presented in this work are based on (Cd,Zn,Mg)Te layers, containing strain free QWs designed and grown to meet the above-postulated properties. After presenting the structure design and the applied technology, we show the results of optical measurements which prove that the conditions for the strong coupling between cavity photons and QWs excitons are achieved already in a single QW structure, and the polariton lasing regime is obtained for a structure embedding 4 QWs containing manganese showing that Mn ions do not destroy the strong coupling regime.

\section{Microcavity structures}

The microcavity structures are grown by molecular beam epitaxy (MBE) on $2^{\prime\prime}$ $(100)$ oriented GaAs substrates. The growth temperature of the buffer and structure was $T_{growth}=346^{\circ} C$. The DBRs are made of Cd$_{0.77}$Zn$_{0.13}$Mg$_{0.10}$Te for the high refractive index layers  and Cd$_{0.43}$Zn$_{0.07}$Mg$_{0.50}$Te for the low refractive index layer. The different Mg concentrations are obtained by using two independent Mg sources \cite{Rousset_JCG2014}. This allows for faster growth processes than growing a digital alloy in the form of a (Cd,Zn)Te|(Cd,Zn,Mg)Te superlattice for the low refractive index layers.\cite{Rou_JCG2013}  The nominal thicknesses of the DBR layers are respectively $d_{high}=61.7$~nm and $d_{low}=69.8$~nm. These thicknesses result in the resonant wavelength $\lambda_0=740$~nm, in resonance with the emission energy of Cd$_{0.84}$Zn$_{0.16}$Te / Cd$_{0.77}$Zn$_{0.13}$Mg$_{0.10}$Te QWs embedded in the cavity. The \emph{in situ} measurement of reflectivity at the desired resonant wavelength during the whole growth process allows for a precise control of the layer thicknesses. During the growth of the DBRs and cavity, the sample was not rotated in order to create a wedge which gives access to exciton-photon detuning in a wide range by selection of the position on the sample (see Figs. \ref{1QW} (a) and \ref{4QW}).

The key point underlying the original design of our structure is the lattice matching to MgTe of the whole microcavity, including the QWs \cite{Rou_JCG2013}. Alloying Cd$_{0.84}$Zn$_{0.16}$Te (lattice matched to MgTe) with MgTe allows one to easily tune the refractive index and energy gap of the layers through the Mg content of the layers, keeping the lattice constant unchanged. In addition, still owing to the design of the structure, the strain induced by small deviations of the Zn content in the DBR layers is negligible: a change of the Zn content by $1\%$ induces a relative mismatch between the DBR layers of $\Delta a / a \approx 0.03\%$ (comparing to $0.12\%$ in the case of well known AlAs/GaAs based DBR), which is crucial for reproducibility of the growth process. Since the QWs embedded in the cavity are strain free, it is possible to grow an arbitrary number of identical QWs in the cavity.

\begin{figure}[!h]
\centering
\includegraphics[width=0.95\linewidth]{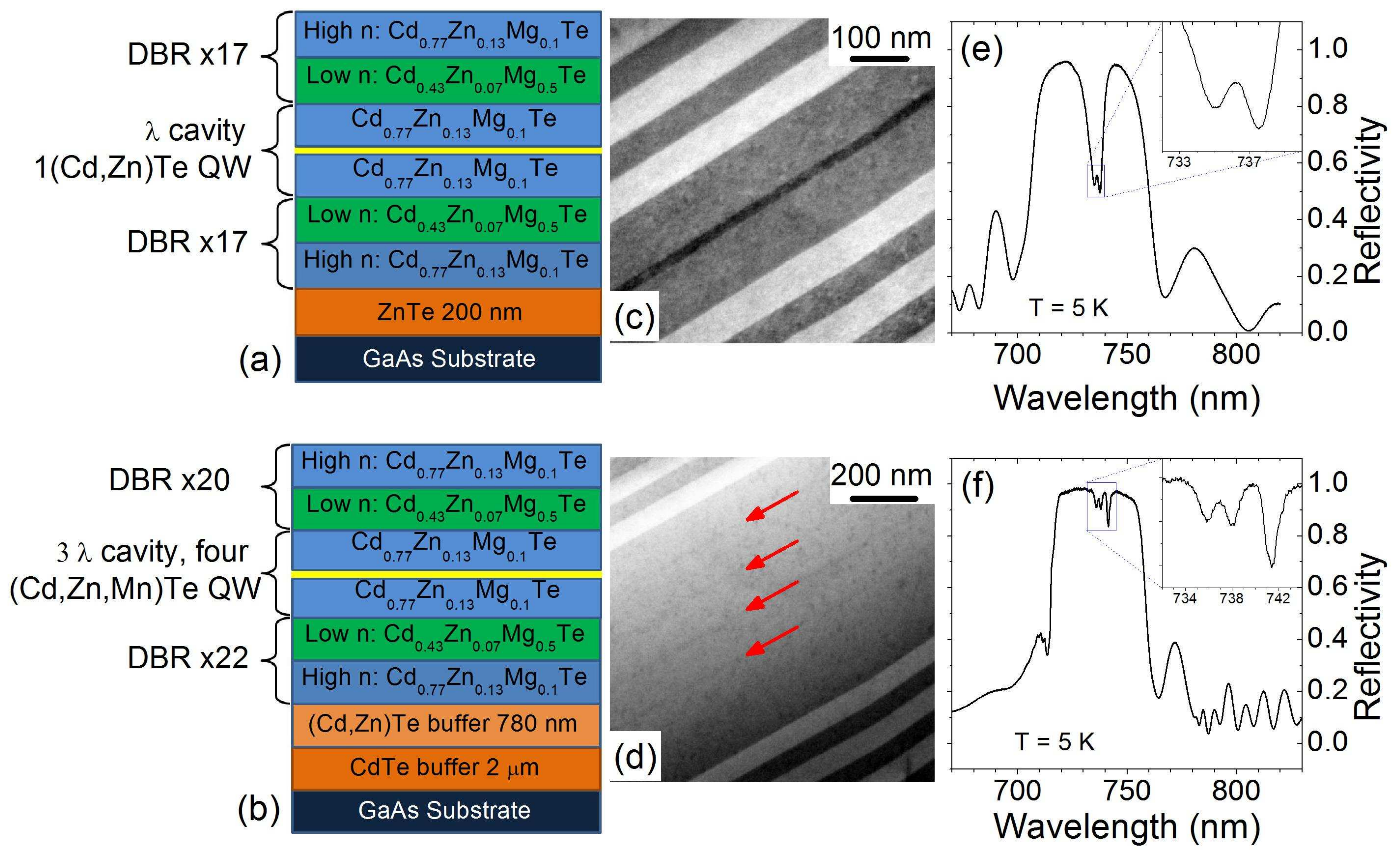}
\caption{(a)-(b) Scheme, (c)-(d) STEM image of the cavity region embedded between the DBRs. The red arrows in (d) figure the positions of the 4 QWs. (e)-(f) reflectivity spectra of the microcavities in the vicinity of the exciton-photon resonance taken at $T = 5 K$. Upper panel (a), (c), (e): structure with a single (Cd,Zn)Te / (Cd,Zn,Mg)Te QW  embedded in a $\lambda$ cavity, lower panel (b), (d), (f): structure with 4 (Cd,Zn,Mn)Te / (Cd,Zn,Mg)Te QWs embedded in a $3\lambda$ cavity. As seen in the reflectivity spectra, at resonance we observe the exciton-photon mode splitting, characteristic for the strong coupling regime.
\label{STEM_ref}}
\end{figure}

The two samples presented in this paper are obtained by the same growth method for the DBRs. As shown in Fig. \ref{STEM_ref} (e) and Fig. \ref{1QW}, the strong coupling regime is obtained with a single Cd$_{0.84}$Zn$_{0.16}$Te / Cd$_{0.77}$Zn$_{0.13}$Mg$_{0.10}$Te QW in a $\lambda$ cavity sandwiched between two 17 fold Bragg reflectors. In order to enhance the quality factor of the cavity and the coupling strength, \cite{Kavokin_MC2007} the second structure consists of 4(Cd,Zn,Mn)Te QWs embedded in a $3\lambda$ cavity sandwiched between a 22 fold (bottom DBR) and 20 fold (top DBR) Bragg reflectors. Additional incorporation of the Mn in the QWs in the second structure allows us to verify that Mn ions do not affect the strong coupling regime and the onset of the polariton lasing regime. As shown in Fig. \ref{STEM_ref} (e) and (f), the optimization of the structure resulted in an enhancement of the quality factor from $\approx100$ for the single QW structure to $\approx 500$ for the cavity embedding 4 QWs.

\section{Strong coupling}

\begin{figure}[!h]
\includegraphics[width=0.8\linewidth]{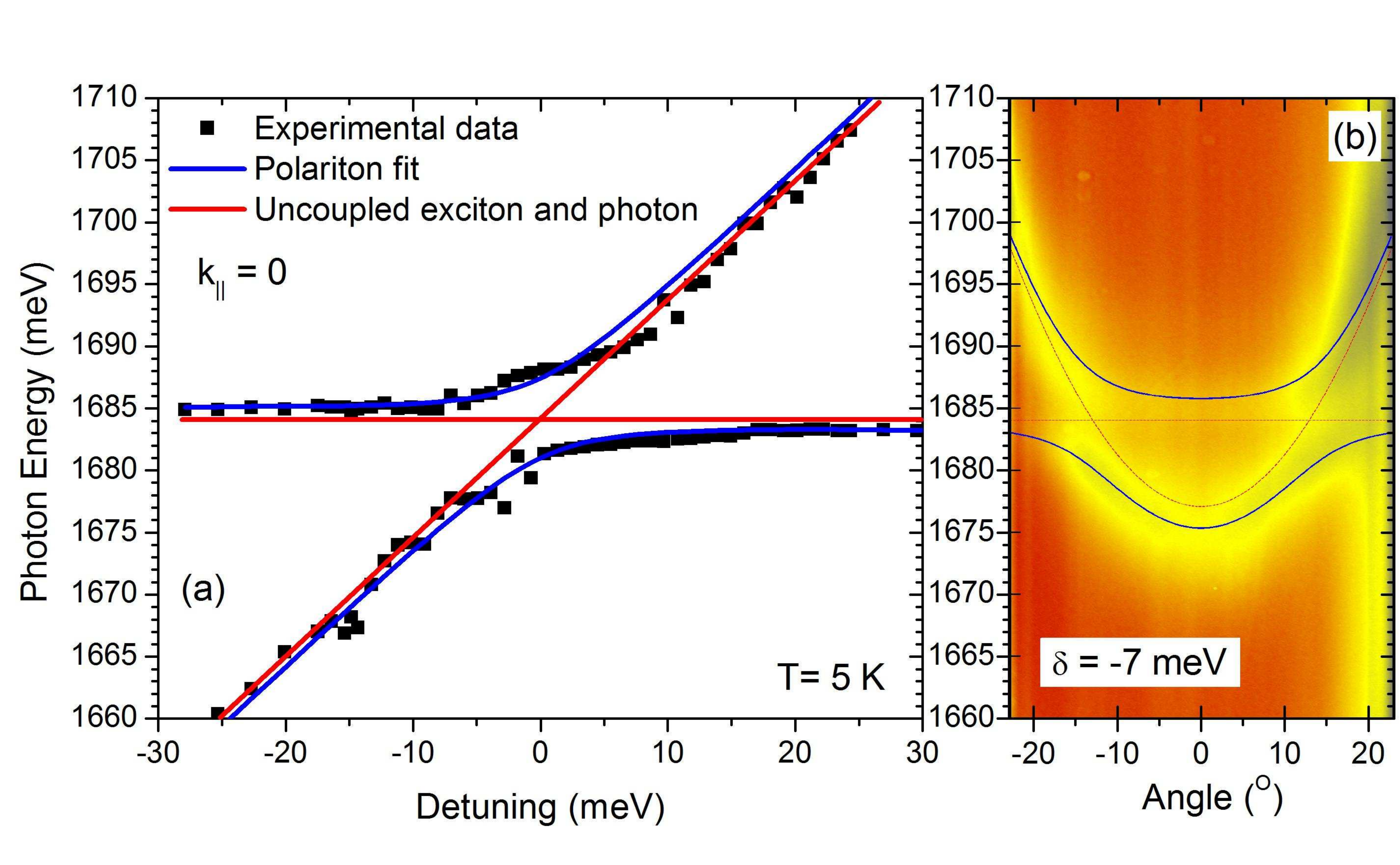}
\caption{(a) Energy positions of the upper and lower polaritons as a function of the exciton-photon detuning for the single QW structure with the anticrossing characteristic for the strong coupling regime. (b) Angle resolved reflectivity spectrum for a detuning $\delta=-7$ meV showing the dispersion of the cavity polaritons. In both cases the Rabi splitting is $\Omega_{1\mathrm{QW}}\approx 6.5$ meV.
\label{1QW}}
\end{figure}

The samples have been characterized by reflectivity measurements in a helium  continuous flow cryostat at temperature $T=5$~K. Taking advantage of the structure wedge resulting from the growth procedure (the sample was not rotated during the MBE growth), different points at the sample give access to various values of the exciton - photon detuning. Plotting the energy positions of the polaritons as a function of the detuning allows us to determine the Rabi splitting characterizing the exciton - photon coupling strength. Fig. \ref{1QW} (a) shows the typical exciton - cavity photon  anticrossing with a Rabi splitting of $\Omega_{1\mathrm{QW}} \approx 6.5$ meV for the single QW microcavity. Angle resolved reflectivity measurement measured in far field gives a direct access to the cavity polaritons dispersion in k-space characteristics, as shown in Fig. \ref{1QW} (b).

\begin{figure}[!h]
\includegraphics[width=0.8\linewidth]{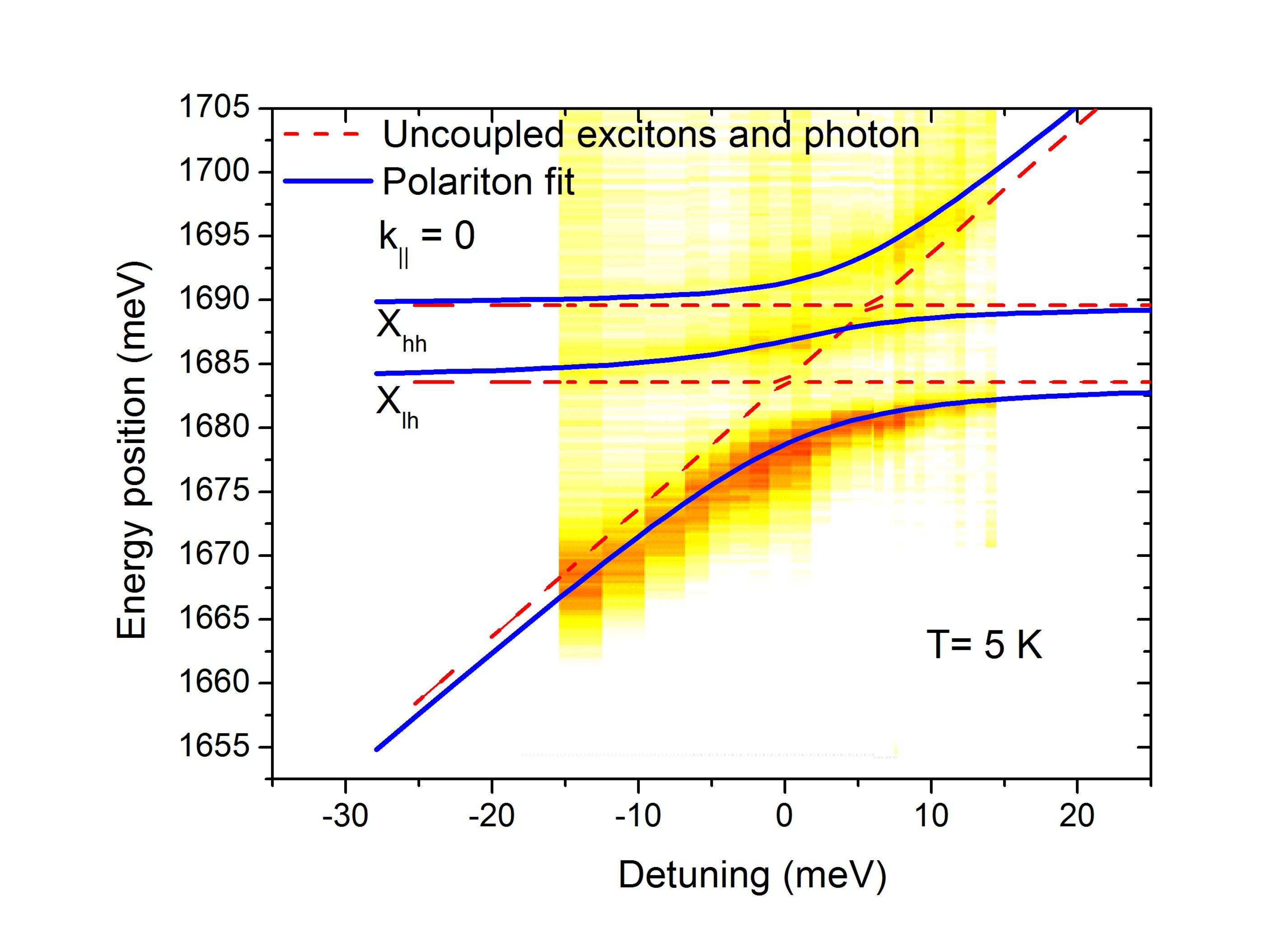}
\caption{Reflectivity map of the 4 QWs structure for various exciton-photon detunings. The three polariton branches evidenced result from the coupling of the cavity mode to heavy hole and  light hole excitons in the quantum wells. The respective coupling strengths (Rabi splitting) to the heavy and the light hole excitons taken as fitting parameters are $\Omega_\mathrm{hh}\approx 9$ meV and $\Omega_\mathrm{lh}\approx 6$ meV.
\label{4QW}}
\end{figure}

The structure containing 4 QWs (Fig. \ref{STEM_ref} (b), (d), (f)) exhibits three polariton branches corresponding to the coupling of the cavity photon to the heavy hole and light hole exciton as shown in Fig. \ref{4QW}. The fit to the experimental data resulted in the determination of the coupling strength (Rabi splitting) of the cavity photon with the heavy hole excitons and the light hole excitons (fitting parameters): respectively  $\Omega_\mathrm{hh}\approx 9$ meV and $\Omega_\mathrm{lh}\approx 6$ meV.

\section{lasing}

\begin{figure}[!h]
\includegraphics[width=1.0\linewidth]{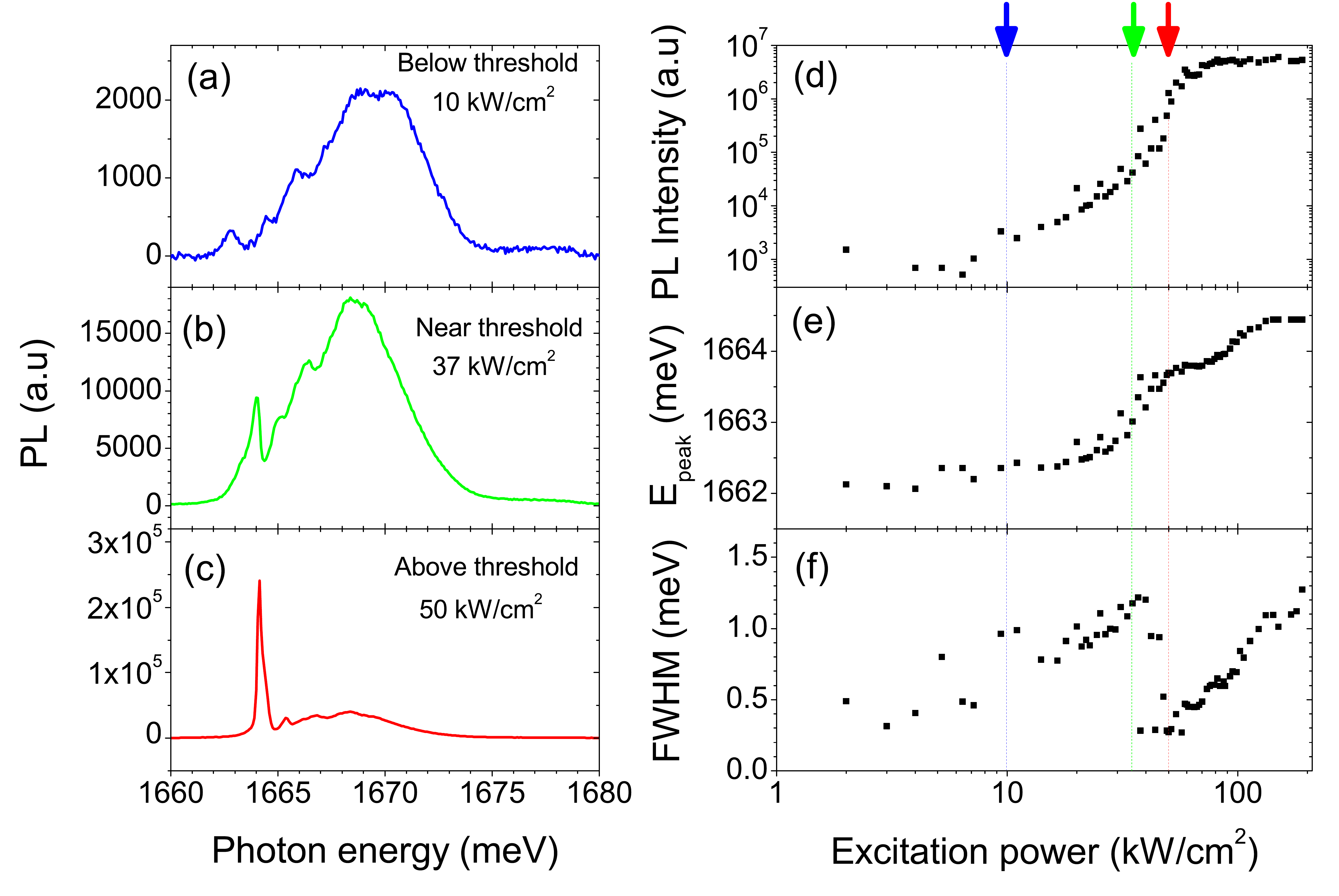}
\caption{PL spectra: (a) below, (b) near and (c) above threshold. Dependence on the excitation power of (d) the intensity, (e) the energy position and (f) the linewidth of the emission line. The arrows mark excitation intensities corresponding to the spectra (a-c). At the threshold power $P_{1} \approx 35\ kW/cm^2$, we observe a blueshift step and narrowing of the emission line which is the fingerprint of the onset of polariton lasing, in the strong coupling regime. The second threshold at $P_{2} \approx 83\ kW/cm^2$, clearly visible in the energy position, indicates the transition to the weak coupling regime and photon lasing.
\label{lasing}}
\end{figure}

Photoluminescence (PL) measurements were performed on the sample containing 4 QWs to investigate the lasing action. The cavity was non resonantly optically pumped by a Ti:Al$_2$O$_3$ laser operating in a femtosecond pulse mode, with energy tuned the first reflectivity minimum on the high energy side of the cavity stopband ($1.76$ eV). Fig. \ref{lasing} (a), (b), and (c) shows the emission of the lower polariton for power excitations respectively below, near and above threshold: at the threshold power $P_{1} \approx 35\ kW/cm^2$ a narrow emission line appears in the spectrum and dominates at higher power excitations. The integrated PL of this emission line exhibits the typical threshold dependence on the excitation power (Fig. \ref{lasing} (d). This threshold is accompanied by a blueshift step due to polariton-polariton interactions and a sudden narrowing of the emission line ((Fig. \ref{lasing} (e), (f)), as a signature of the increased coherence in the system. These observations are in agreement with the onset of the polariton lasing regime.\cite{Bajoni_PRL2008, Kamman_NJP2012} The second threshold at excitation power $P_{2} \approx 83\ kW/cm^2$ visible in the energy position of the emission line, we associate to the vanishing of the strong coupling regime due to the high exciton population and the onset of the stimulated emission amplification \cite{Bajoni_PRB2007, Kamman_NJP2012} (photon lasing regime).

\section{Discussion}

The results presented above show that the conditions for polariton lasing are reached in the Te-based microcavity structures we have designed. The Rabi splitting obtained for the single QW structure ($\Omega_{1\mathrm{QW}}\approx 6.5$ meV) is comparable to the one observed in the previous Te-based microcavity structures \cite{Ulmer_SLandMicrostruc97,Brunetti_SLandMicrostruc2007} or Se-based microcavities. \cite{Sebald_APL2012,Klein_APL2015,Klembt_PRL2015} However, the original structure  designed and studied here, as a result of matching its lattice constant to MgTe, additionally provides a high reproducibility, a low sensitivity to strain that might be induced by Zn concentration fluctuations in the deposited layers and an easy tuning of the optical properties making it very flexible and easy to operate. In addition, the absence of magnetic elements in the DBRs provides an additional degree of freedom given by the possibility to implement Mn precisely in the desired element of the structure (e.g. in the QWs). This will enable magneto-optical studies of cavity polaritons, in which the excitonic part exhibits enhanced magneto-optical effects. In that sense, the polariton lasing regime observed in the structure containing 4 QWs doped with Mn is the first milestone towards such investigations. In a larger perspective, such a microcavity structure is well suited for embedding\cite{Pac_CGandD2014,Jomek_ACSnano2014} CdTe quantum dots (QD) with single magnetic impurities. \cite{Besombes_PRL2004, Kobak_NComms2014, Besombes_PRB2014, Goryca_PRL2014} This will be of great interest in the frame of investigations on refined effects, the development of solotronics \cite{Koen_NatureMat2011, Kobak_NComms2014} and a preliminary step towards the strong coupling between a cavity photon and exciton in a single QD doped with a single magnetic impurity.\cite{Andrade_PRB2012}

\section{Conclusion}

In this work we have demonstrated for the Te based microcavities containing unstrained (Cd,Zn)Te quantum wells the achievement of the strong coupling regime for a single QW microcavity structure and we evidence the polariton lasing regime in a structure embedding 4 QWs containing magnetic ions. The material system proposed in this work combines several advantages directly resulting from its conception: (i) A high reproducibility of the growth process. (ii) A low influence on strain related to Zn content fluctuations from one growth process to another. Both as a result of the lattice matching of (Zn, Cd, Mg)Te based layers to MgTe constituting the DBRs. (iii) The easy tuning of the optical properties through the Mg content of the DBR layers keeping the lattice constant unchanged. (iv) The possibility of precise placement of magnetic ions in the structure free of magnetic elements enabling the study of \emph{semimagnetic} polaritons.

\begin{acknowledgments}
This work was supported by the Polish Ministry of Science and Higher Education in years 2015--2017 as research grants ''Iuventus Plus'' (projects IP2014 040473 and IP2014 034573), NCN projects DEC-2014/13/N/ST3/03763, DEC-2013/10/E/ST3/00215, DEC-2013/09/B/ST3/02603, DEC-2013/09/D/ST3/03768, DEC-2011/02/A/ST3/00131, DEC-2011/01/D/ST7/04088, NCBiR project "Lider" and Polish Foundation for Science (FNP) subsidy ''Mistrz''. The research leading to these results has received funding from the European Union
Seventh Framework Programme (FP7/2007-2013) under grant agreement No.316244.
\end{acknowledgments}

\end{document}